\documentclass[aps,prb,twocolumn,showpacs,showkeys,preprintnumbers,groupedaddress,amsmath,amssymb,]{revtex4-1}

\usepackage{graphicx}
\usepackage{dcolumn}
\usepackage{bm}
\usepackage{hyperref}
\usepackage[dvips]{color}
\usepackage{booktabs}
\usepackage{longtable}
\usepackage{array}
\usepackage{dcolumn,booktabs}

\begin{document}


\title{Electrical conduction processes in ZnO in a wide temperature range 20--500 K}

\author{Chien-Chi Lien$^1$, Chih-Yuan Wu$^2$}
\email[Electronic address: ]{016287@mail.fju.edu.tw}

\author{Zhi-Qing Li$^3$}
\author{Juhn-Jong Lin$^{1,4,}$}
\email[Electronic address: ]{jjlin@mail.nctu.edu.tw}

\affiliation{$^1$Institute of Physics, National Chiao Tung University, Hsinchu 30010, Taiwan}

\affiliation{$^2$Department of Physics, Fu Jen Catholic University, New Taipei City 24205, Taiwan}

\affiliation{$^3$Tianjin Key Laboratory of Low Dimensional Materials Physics
and Preparing Technology, Faculty of Science, Tianjin University, Tianjin 300072, China}

\affiliation{$^4$Department of Electrophysics, National Chiao Tung University, Hsinchu 30010, Taiwan}

\date{\today}

\begin{abstract}

We have investigated the electrical conduction processes in as-grown and
thermally cycled ZnO single crystal as well as as-grown ZnO polycrystalline
films over the wide temperature range 20--500 K. In the case of ZnO single
crystal between 110 and 500 K, two types of thermal activation conduction
processes are observed. This is explained in terms of the existence of both
shallow donors and intermediately deep donors which are consecutively excited
to the conduction band as the temperature increases. By measuring the
resistivity $\rho(T)$ of a given single crystal after repeated thermal
cycling in vacuum, we demonstrate that oxygen vacancies play an important
role in governing the shallow donor concentrations but leave the activation
energy ($\simeq$27$\pm$2 meV) largely intact. In the case of polycrystalline
films, two types of thermal activation conduction processes are also observed
between $\sim$150 and 500 K. Below $\sim$150 K, we found an additional
conduction process due to the nearest-neighbor-hopping conduction mechanism
which takes place in the shallow impurity band. As the temperatures further
decreases below $\sim$80 K, a crossover to the Mott variable-range-hopping
conduction process is observed. Taken together with our previous measurements
on $\rho (T)$ of ZnO polycrystalline films in the temperature range 2--100 K
[Y. L. Huang {\it et al.}, J. Appl. Phys. \textbf{107}, 063715 (2010)], this
work establishes a quite complete picture of the overall electrical conduction
mechanisms in the ZnO material from liquid-helium temperatures up to 500 K.

\end{abstract}


\maketitle

\section{Introduction}

Zinc oxide (ZnO) is a wide band gap semiconductor with a direct energy gap of
$\approx$3.4 eV at room temperature. This class of material is usually
unintentionally (or, natively) doped with n-type impurities, such as oxygen
vacancies, Zn interstitials, and impurity hydrogen atoms. However, the exact
doping behavior of different types of unintentional and intentional (e.g.,
group I or group III) dopants is still under much theoretical and experimental
debate. \cite{Janotti09,McCluskey09,Ozgur05}

ZnO has recently attracted intense attention owing to its potential prospects
in optoelectronic applications. \cite{Ellmer08,LiZQ06,Look01} Compared with
the widely studied optical properties, the electrical conduction processes in
ZnO materials (single crystals, polycrystalline films, etc.) have not much
been investigated over a wide range of temperature. \cite{Kumar08, Tiwari04,
Heluani06, Tsurumi99} Physically, the measurements on the temperature
dependence of resistivity, $\rho (T)$, can provide very useful information on
the carrier transport mechanisms as well as the associated impurity levels and
their distributions in energy in a given semiconducting material. Recently,
Huang {\it et al.} \cite{Huang10} have investigated the electrical conduction
processes in a series of oxygen deficient polycrystalline ZnO films
($\approx$1 $\mu$m thick) in the temperature range 2--100 K. They observed the
three-dimensional Mott variable-range-hopping (VRH) conduction process below
$\sim$100 K. As the temperature further decreased below $\sim$25 K, a
crossover to the Efros-Shklovskii VRH conduction process was found. These two
types of VRH conduction processes in ZnO films have been successfully
explained in a coherent manner. On the other hand, the charge transport
properties above 100 K has not been explicitly addressed in that work, due to
the lacking of $\rho (T)$ data above 300 K. In this work, we aim at extending
the measurements on $\rho (T)$ to temperatures up to 500 K in order to further
unravel the overall electrical conduction mechanisms in ZnO, especially at
$T$$\gtrsim$100 K. We have measured an as-grown ZnO single crystal and
remeasured the series of as-grown ZnO polycrystalline films which were first
studied in Ref. \onlinecite{Huang10}. The as-grown single crystal has been
{\em repeatedly} thermally cycled {\em in vacuum} and their $\rho (T)$ has been measured
during each thermal cycling process. The polycrystalline films had been deposited in
different oxygen atmospheres to contain different amounts of oxygen vacancies
so that they possessed differing shallow donor concentrations when as
prepared. Our new results are reported below.

We would like to emphasize that all the $\rho(T)$ measurements reported in this study have been carried out in a vacuum ($\lesssim$$10^{-4}$ torr) and in a ``dark" environment without involving any photo-induced carriers. Therefore, complex thermal annealing effects which might result from resistivity measurements performed in, e.g., an air (or an active gas) atmosphere at highly elevated temperatures, \cite{Zhang04} can be safely avoided.

This paper is organized as follows. In Sec. II, we briefly discuss our
experimental method for sample fabrication and resistance measurements. Our
results of the temperature dependence of resistance in the wide temperature
interval 20--500 K and their physical interpretations are presented in Sec.
III. Section IV contains our conclusion as well as a summary of the overall
electrical conduction mechanisms in the ZnO materials.

\section{Experimental Method}

Our ZnO single crystal (2.52$\times$1.94$\times$0.45 mm$^3$) were grown by the
hydrothermal method and was obtained from a commercial supplier.
\cite{supplier} Unfortunately, the detailed growth conditions were
unavailable. However, the $\rho (T)$ data do provide us meaningful information
on the carrier transport processes in the ZnO material (see below). This
as-grown crystal has a room temperature resistivity of
$\approx$5.45$\times$$10^4$ $\Omega$ cm. Our ZnO polycrystalline films were
fabricated by the standard rf sputtering deposition method on glass
substrates. The starting ZnO target was of 99.99\% purity, and the glass
substrates were held at 550$^\circ$C during the deposition process. The films
were deposited in varying mixtures of argon and oxygen gases. The $\rho (T)$
of these films had previously been measured and discussed by Huang {\it et
al.} \cite{Huang10} for $T$$\lesssim$100 K.

In this work, four-probe resistance measurements were carried out using a
Keithley Model K-6430 as a current source and a high-impedance (T$\Omega$)
Keithley Model K-617 as a voltmeter. The samples were mounted on the sample
holder which was situated inside a stainless vacuum can in a JANIS CCS350 closed-cycle
refrigerator (10--500 K). The working pressure in the vacuum can was $\lesssim$$10^{-4}$ torr throughout all resistivity measurement processes, whether the vacuum can was connected to a diffusion pump (at $T$$\gtrsim$250 K) or not (at $T$$\lesssim$250 K, see below for further discussion). Moreover, the stainless vacuum can contained no optical windows, and thus our measurements were performed in the dark, i.e., this experiment did not invoke any kind of photo-induced currents. The four-terminal current leads and voltage leads
(pogo pins) were attached to the samples with the aid of silver paste. It
should be noted that the resistances reported in this work were all measured
by scanning the current-voltage ($\textit{I-V}$) characteristics at various
fixed temperatures between 20 and 500 K. The resistance at a given temperature
was determined from the regime around the zero bias voltage, where the
$\textit{I-V}$ curve was linear. In fact, our $\textit{I-V}$ curves in every
sample were linear over a wide range of bias voltage. The inset of
Fig.~\ref{Fig.1} shows the linear $I$-$V$ characteristics of the as-grown ZnO
single crystal at four representative temperatures, as indicated in the
caption to Fig.~\ref{Fig.1}.

\section{Results and discussion}
\subsection{ZnO single crystal: thermal activation conduction over 110--500 K}

\begin{figure}[htbp]
\begin{center}
\includegraphics[scale=0.35]{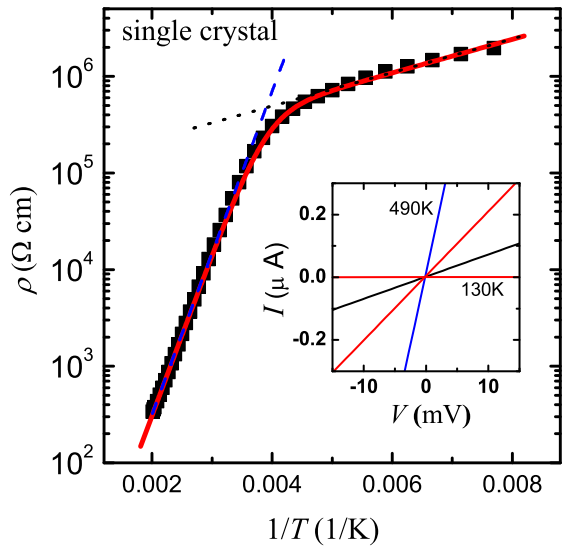}
\caption{(color online) Logarithm of resistivity as a function of reciprocal
temperature for as-grown ZnO single crystal between 110 and 500 K. The symbols
are the experimental data and the solid curve is a least-squares fit to
Eq.~(\ref{Eq.(1)}). The straight dashed and dotted lines are guides to the
eye. Inset: Current-voltage curves at four temperatures of 490,
410, 370, and 130 K. Notice that the $I$-$V$ characteristics are linear, i.e.,
ohmic.} \label{Fig.1}
\end{center}
\end{figure}

Figure~\ref{Fig.1} plots the measured resistivity as a function of reciprocal
temperature for the as-grown single crystal between 110 and 500 K. (The
resistance below $\sim$110 K became too large to be accurately measured.) Two
distinct slopes are explicitly seen in the two different temperature regimes
of $\sim$110--220 K (straight dotted line) and $\sim$290--500 K (straight
dashed line), respectively. Such Arrhenius-type behavior immediately suggests
that the responsible charge transport arises from the thermal activation
conduction processes. Indeed, we found that our measured $\rho (T)$ data can
be quantitatively described by the equation
\begin{equation}\label{Eq.(1)}
\rho(T)^{-1} = \rho_1^{-1}\, e^{-E_1/k_BT} + \rho_2^{-1}\, e^{-E_2/k_BT} \,,
\end{equation}
where $\rho_1$ and $\rho_2$ are temperature insensitive resistivity
prefactors, and $E_1$ and $E_2$ are the relevant activation energies
associated with the two kinds of thermal activation conduction processes.
$k_B$ is the Boltzmann constant. In Fig. 1, the solid curve is a least-square
fit to Eq.~(\ref{Eq.(1)}). It clearly manifests that Eq.~(\ref{Eq.(1)}) can
well describe the experimental result. Our fitted values of $\rho_1$, $E_1$,
$\rho_2$, and $E_2$ are listed in Table I. This observation illustrates that
there exist in the as-grown single crystal a group of shallow donors with an
activation energy of $E_2$$\approx$29 meV and a group of intermediately
deep donors with an activation energy of $E_1$$\approx$330 meV. (In this work,
we shall maintain the convention $E_1$$>$$E_2$.) At the sufficiently high
measurement temperatures of $T$$\gtrsim$290 K, both the shallow and the
intermediately deep donors are excited to the conduction band and are
responsible for the electrical transport. However, the exponential temperature
dependence of $\rho$ is mainly governed by the number of those intermediately
deep donors which are being excited to the conduction band (the so-called
$E_1$-conduction channel). As $T$ reduces below $\sim$220 K, the
intermediately deep donors become essentially intact while the shallow donors
can still be readily excited to the conduction band. The $T$ dependence of
$\rho$ is thus largely determined by the number of those shallow donors being
excited to the conduction band (the so-called $E_2$-conduction channel). In
short, Eq.~(\ref{Eq.(1)}) represents the ``band conduction" processes.

\begin{table*}
\caption{\label{Table I} Relevant parameters of as-grown and thermally cycled ZnO single crystal. $\rho_{1}$, $E_{1}$, $\rho_{2}$, and $E_{2}$ are
defined in Eq.~(1).}
\begin{ruledtabular}
\tabcolsep=12pt
\begin{tabular}{lccccc}

single crystal & $\rho$(300\,K) ($\Omega$ cm) & $\rho_{1}$ ($\Omega$ cm) &
$E_{1}$ (meV) & $\rho_{2}$ ($\Omega$ cm) & $E_{2}$ (meV)\\\hline

as-grown & 5.45$\times10^{4}$ & 0.144 & 334 & 1.47$\times10^{5}$ & 29.2\\

first-cycled & 4.56$\times10^{4}$ & 0.154 & 330 & 2.50$\times10^{4}$ & 28.1\\

second-cycled & 2.70$\times10^{4}$ & 0.145 & 332 & 1.72$\times10^{4}$ & 24.7\\

third-cycled & 2.01$\times10^{4}$ & 0.125 & 337 & 1.02$\times10^{4}$ & 26.4\\

\end{tabular}
\end{ruledtabular}
\end{table*}

It should be noted that our extracted $E_1$ and $E_2$ values are in good
accord with the corresponding activation energies previously obtained by Look
\cite{Look01} ($\sim$340 meV), Wenckstern {\it et al.} \cite{Wenckstern07}
($\sim$34--37 and $\sim$300--370 meV), and Schifano {\it et al.}
\cite{Schifano09} ($\sim$30, $\sim$50, and $\sim$290 meV) in various ZnO
single crystals and in different temperature intervals. Recently, in a series
of single-crystalline ZnO nanowires, Chiu {\it et al.} \cite{Chiu09} (Tsai
{\it et al.} \cite{Tsai10}) have also reported an $E_2$ activation energy of
$\sim$30--40 ($\sim$25) meV. In other words, we may conclude that different
ZnO single crystals grown by differing methods and under differing conditions
possess essentially similar types of shallow and intermediately deep donor
levels. (Obviously, the donor concentrations would vary from experiment to
experiment.)

\begin{figure}[htbp]
\begin{center}
\includegraphics[scale=0.35]{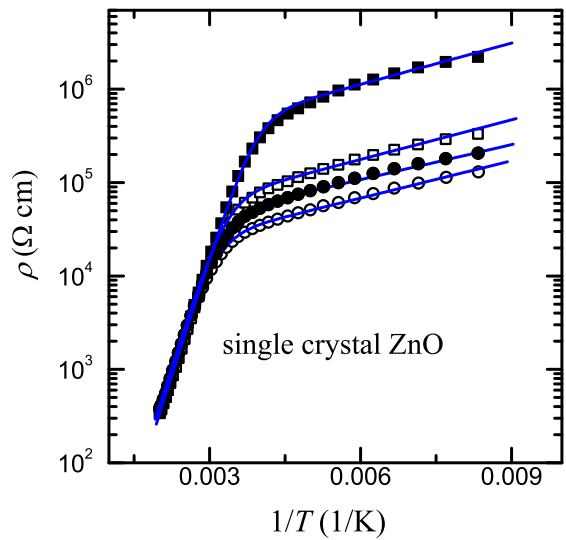}
\caption{(color online) Logarithm of resistivity as a function of reciprocal
temperature for the as-grown (closed squares, taken from Fig.~\ref{Fig.1}),
first-cycled (open squares), second-cycled (closed circles), and
third-cycled (open circles) ZnO single crystal. The solid curves are
least-squares fits to Eq.~(\ref{Eq.(1)}), see text.} \label{Fig.2}
\end{center}
\end{figure}

In Fig.~\ref{Fig.2}, we plot our measured resistivities as a function of
reciprocal temperature for the as-grown and then repeatedly thermally
cycled single crystal. The closed squares (taken from Fig.~\ref{Fig.1}),
open squares, closed circles, and open circles represent the as-grown,
first-cycled, second-cycled, and third-cycled samples, respectively.
In practice, our $\rho (T)$ data for every sample were measured from 500 K
down. The sample was placed in the dark vacuum can and first heated to 500 K. The
$T$ dependence of $\rho$ was then measured with progressive lowing of
temperature. During the data acquisition process from 500 K down to 250 K, the
dark vacuum can was connected to a diffusion pump for continuous pumping. The
diffusion pump was disconnected when the temperature reached 250 K. \cite{backflow} Such a
measurement process (which lasted for $\sim$270 min between 250 and 500 K) had
effectively served as a ``thermal annealing" (in vacuum) process. Therefore, the oxygen
contain in the sample was reduced from that in the previous run. In other
words, the amount of oxygen vacancies increased, leading to a decrease in the
sample resistivity. We have repeated the $\rho (T)$ measurements in the
temperature interval 110--500 K for four times, and denoted the sample by
as-grown, first-cycled, second-cycled, and third-cycled samples, as mentioned.

Inspection of Fig.~\ref{Fig.2} indicates that, in the temperature interval
$\sim$330--500 K, the $\rho (T)$ data collapse closely onto a straight line.
On the other hand, below $\sim$330 K, the measured $\rho (T)$ data
systematically decrease after each repeated thermal cycling. Quantitatively, the
resistivity ratios of the as-grown sample to the third-cycled sample are
$\simeq$1.0 at 460 K, $\simeq$2.7 at 300 K, and $\simeq$11 at 220 K. This
variation in resistivity ratio suggests that the rich amount of the
intermediately deep donors \cite{rich} and their associated activation energies, i.e., the
$E_1$ values, are barely affected by our measuring and/or thermal cycling process.
This is expected, since the activation energies associated with these donor
levels are comparatively large while our thermal cycling temperature is relatively low.

What is more interesting is the $T$ behavior of $\rho$ between 110 and 220 K.
While the $\rho$ value at a given $T$ decreases with repeated thermal
cycling as mentioned, the {\em temperature dependence} of $\rho$ maintains
essentially {\em unchanged}. Numerically, our least-squares fitted slopes,
i.e., the $E_2$ values, in this temperature interval vary only by $\sim$10\%, see
Table I. This quantitative result is meaningful. It strongly implies that the
amount of oxygen vacancies, namely, the shallow donor concentration, in the
sample is increased after each repeated thermal cycling. Nevertheless, the thermal
activation energy, namely, the shallow donor level below the conduction band
minimum, remains essentially unchanged. (See Ref. \onlinecite{disperse} for a brief discussion on the thermal-cycling-induced weak dispersion of the shallow donor energy levels.) This observation is strongly supportive of the important roles of oxygen vacancies as the shallow
donors in ZnO. On the theoretical side, based on the formation energy
consideration, it has recently been argued that oxygen vacancies would form
negative-U centers, rather than shallow donors, in the ZnO material.
\cite{Janotti05,Vlasenko05,Zhang01,Lany05} This theoretical prediction is not
supported by the present result, Fig.~\ref{Fig.2}. This puzzling issue requires further clarification.

\subsection{ZnO polycrystalline films}
\subsubsection{Thermal activation conduction and nearest-neighbor-hopping conduction: 90--500 K}

\begin{figure}[htbp]
\begin{center}
\includegraphics[scale=0.35]{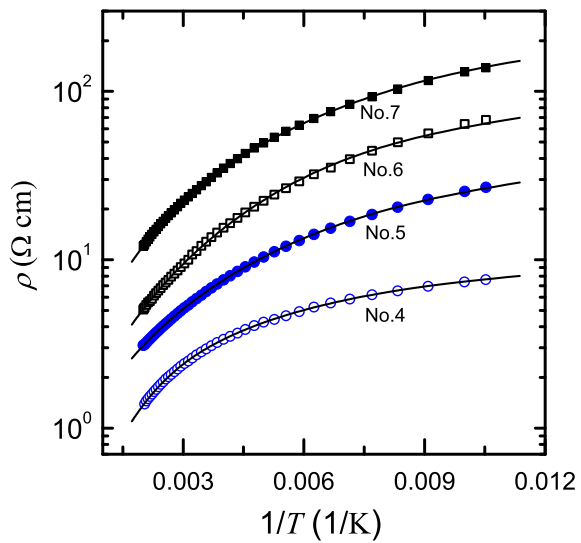}
\caption{Logarithm of resistivity as a function of reciprocal temperature for
four ZnO polycrystalline films between 90 and 500 K, as indicated. The symbols
are the experimental data and the solid curves are least-squares fits to
Eq.~(\ref{Eq.(2)}).} \label{Fig.3}
\end{center}
\end{figure}

Figure~\ref{Fig.3} shows the variation in the logarithm of resistivity with
reciprocal temperature for four ZnO polycrystalline films between 90 and 500
K, as indicated. In sharp contrast to the case of single crystal
(Fig.~\ref{Fig.2}), Fig.~\ref{Fig.3} clearly reveals that $\rho$ smoothly
increases with decreasing temperature. That is, there does not exist a visible
straight regime in any temperature interval in the log$\rho$--(1/$T$) plot
from 500 down to 90 K. Quantitatively, we found that these results can not be
described by Eq.~(\ref{Eq.(1)}). Instead, they can be described by the
following equation
\begin{equation}\label{Eq.(2)}
\rho(T)^{-1} = \rho_1^{-1}\, e^{-E_1/k_BT} + \rho_2^{-1} e^{-E_2/k_BT} +
\rho_3^{-1}\, e^{-E_3/k_BT} \,,
\end{equation}
where $\rho_1$, $E_1$, $\rho_2$, and $E_2$ have the similar meaning as defined
in Eq.~(\ref{Eq.(1)}). The third term on the right hand side represents a new,
additional electrical conduction channel which is characterized by a
temperature insensitive resistivity prefactor $\rho_3$ and a thermal
activation energy $E_3$. (In the following discussion, we shall maintain the
convention $E_1$$>$$E_2$$>$$E_3$.)

Our fitted results with Eq.~(\ref{Eq.(2)}) are plotted as the solid curves in
Fig.~\ref{Fig.3} and the values of the adjustable parameters are listed in
Table II. We obtain the thermal activation energies $E_1$$\approx$113--135
meV, $E_2$$\approx$26--39 meV, and $E_3$$\approx$2.3--5.7 meV. The three types
of conduction processes dominate the $T$ dependence of $\rho$ in the
temperature intervals of $\sim$400--500 K (the $E_1$-conduction channel),
$\sim$150--400 K (the $E_2$-conduction channel), and $\sim$90--150 K (the
$E_3$-conduction channel). Physically, the $E_1$- and the $E_2$-conduction
processes are just those described in the above subsection. However, the $E_1$
activation energy is markedly reduced from $\approx$330 meV for single crystal
to $\approx$125$\pm$10 mev for polycrystalline films (see further discussion
below). The additional $E_3$-conduction channel can be ascribed to the
nearest-neighbor-hopping (NNH) conduction mechanism which takes place in the
shallow ``impurity band." That is, as the temperature decreases below
$\sim$150 K, even the shallow donors (not mentioning the intermediately deep
donors) can no longer be effectively excited to the conduction band. The
charge transport is then largely governed by those electrons which hop from an
occupied state (which lies below the Fermi energy) to an unoccupied
nearest-neighbor state (which lies above the Fermi energy) in the impurity
band. Such a hopping process is phonon assisted and requires a small amount of
energy, i.e., the $E_3$ thermal activation energy. It should be noted that a shallow impurity
band can form in the oxygen deficient ZnO polycrystalline films but not in the
ZnO single crystal, because the former samples contain far higher amounts of
oxygen vacancies. A large amount of oxygen vacancies causes a notable dispersion
of the donor levels or ionization energies, and thus forming an impurity band.
\cite{Mott79} Due to the slight self-compensation characteristic of
unintentionally doped ZnO, \cite{Chiu09,Look03} the impurity band is partially
filled, making the NNH conduction process possible. Based on this picture, the
$E_3$ value should decrease with increasing oxygen vacancies, because there
would be more randomly distributed donor levels per unit energy per unit volume. Indeed,
inspection of Table II reveals that our extracted $E_3$ value (on average)
decreases with the decreasing O$_2$ flux applied during the fabrication process.

\begin{table*}
\caption{\label{Table II} Values of relevant parameters for six oxygen
deficient ZnO polycrystalline films. $\rho_{i}$ ($i$=1, 2, 3) and $E_{i}$ are
defined in Eq. (2). Notice that these samples are taken from Ref.
\onlinecite{Huang10} and remeasured from 500 K down. As a result of thermal
cycling/annealing in this study (see text), the resistivity values are lowered by
$\sim$20 $\%$ from those corresponding values originally reported in Ref.
\onlinecite{Huang10}.}
\begin{ruledtabular}
\tabcolsep=12pt
\newcolumntype{a}[1]{D{-}{-}{#1}}
\newcolumntype{b}[1]{D{.}{,}{#1}}
\begin{tabular}{ccccccccc}
Film & O$_{2}$ flux & $\rho$(300\,K)& $\rho_{1}$ & $E_{1}$ & $\rho_{2}$ &
$E_{2}$
& $\rho_{3}$ & $E_{3}$\\

No. & (SCCM)&($\Omega$ cm)&($\Omega$ cm)&(meV)&($\Omega$ cm)&(meV)&($\Omega$
cm)&(meV)\\ \hline
2 & 0.02 & 1.66 & 0.079 & 120 & 1.89 & 35.5 & 2.63 & 2.3\\
3 & 0.10 & 2.72 & 0.407 & 113 & 0.758 & 39.2 & 5.00 & 3.1\\
4 & 0.15 & 2.01 & 0.189 & 135 & 1.30 & 37.5 & 5.88 & 2.8\\
5 & 0.25 & 5.89 & 0.420 & 135 & 2.70& 26.3 & 370 & 5.7\\
6 & 0.50 & 11.3 & 0.588 & 129 & 4.79 & 27.0 & 333 & 5.6\\
7 & 0.80 & 26.0 & 2.04 &  121 & 9.09 & 37.7 & 100 & 4.3\\
\end{tabular}
\end{ruledtabular}
\end{table*}

In contrast, in the case of single crystal, the shallow donor concentration is
sufficiently low and thus, on average, the donor impurities should lie
sufficiently far apart in space and in energy. Therefore, the NNH conduction
process can only play a negligible role, as compared to the $E_2$-conduction
process, even at temperatures down to 110 K. Inspection of Tables I and II
indicates that the $\rho$(300\,K) values of our polycrystalline films are more
than 3 orders of magnitude lower than that in the single crystal. Such huge
differences in resistivity provide a direct indication of the presence of far
more numerous amounts of oxygen vacancies in polycrystalline films than in
single crystal. (The shallow donor concentrations in the polycrystalline films will be
estimated in the subsection IIIB.3.)

Our extracted $E_2$ values of $\approx$26--39 meV for the shallow donors are
in consistency with those previously reported for the ZnO materials.
\cite{Chiu09,Look03} Our extracted $E_1$ values of $\approx$125$\pm$10 meV are
in line with that ($\approx$110 meV) deduced by Tampo {\it et al.}
\cite{Tampo04} in ZnO films grown by radical source molecular-beam epitaxy.
However, these $E_1$ values are significantly lower than that ($\approx$330
meV) found in the single crystal discussed in the above subsection. The reason
why requires further investigation. One plausible explanation would be that,
as compared with the single crystal, there are additional impurity levels
which were unintentionally introduced during the growth of the polycrystalline
films. On the other hand, since the $E_2$-conduction channel dominates the
electrical-transport behavior up to $\sim$400 K in our films, electrical
measurements up to $T$ sufficiently higher than 500 K would be highly
desirable for an unambiguous determination of the values of the $E_1$
activation energy and the associated charge conduction process.
\cite{thermionic}

\subsubsection{Three-dimensional Mott variable-range-hopping conduction: 20--80 K}

\begin{table*}
\caption{\label{Table III} Values of relevant parameters for six oxygen
deficient ZnO polycrystalline films. $\rho_{M_0}$ and $T_M$ are defined in
Eq.~(\ref{Eq.(3)}). $N(E_F)$ is the electronic density of states at the Fermi
energy, $\bar{R}_{\rm hop,Mott}$ is the average hopping distance, $\xi$ is the
Bohr radius of the shallow donors, and $\bar{W}_{\rm hop,Mott}$ is the average
hopping energy. The values of $\bar{W}_{\rm hop,Mott}$ were calculated for a
representative temperature of 40 K.}
\begin{ruledtabular}
\tabcolsep=12pt
\begin{tabular}{ccrccc}
Film & $\rho_{M_0}$ & $T_{M}$ & $N(E_F)$ &
{\Large $\frac{\bar{R}_{\rm hop,Mott}}{\xi}$} & $\bar{W}_{\rm hop,Mott}$\\
No. & ($\Omega$ cm)&(K)&(J$^{-1}$ m$^{-3}$)&  &(meV)\\ \hline
2   &   1.36    &   65      &   2.5$\times10^{48}$  &   1.07/$T^{1/4}$  &   0.98\\
3   &   0.61    &   3510    &   4.6$\times10^{46}$  &   2.89/$T^{1/4}$  &   2.64\\
4   &   1.41    &   854     &   1.9$\times10^{47}$  &   2.03/$T^{1/4}$  &   1.85\\
5   &   1.15    &   10400   &   1.6$\times10^{46}$  &   3.79/$T^{1/4}$  &   3.35\\
6   &   1.36    &   23600   &   6.9$\times10^{45}$  &   4.65/$T^{1/4}$  &   4.14\\
7   &   4.18    &   15200   &   1.1$\times10^{46}$  &   4.16/$T^{1/4}$  &   3.68\\

\end{tabular}
\end{ruledtabular}
\end{table*}

As mentioned, our ZnO polycrystalline films had been measured previously in
Ref. \onlinecite{Huang10}. In this work, we have extended the $\rho (T)$
measurements on these films up to 500 K. Notice that these measurements have
led to thermal cycling/annealing effect (in vacuum) on these films. As a result, their $\rho (T)$
values are changed from the corresponding values originally reported in Ref.
\onlinecite{Huang10}. In particular, the $\rho (T)$ values are largely reduced
in those films grown under high oxygen atmospheres. For instance, the
$\rho$(300\,K) value decreased from 206 to 26 $\Omega$ cm in the film No. 7.
Nevertheless, we would expect the charge conduction mechanisms, i.e., the Mott
and the Efros-Shklovskii VRH conduction processes previously observed below
$\sim$90 K as mentioned in the Introduction, to be robust. This assertion is
examined in the following.

\begin{figure}[htbp]
\begin{center}
\includegraphics[scale=0.35]{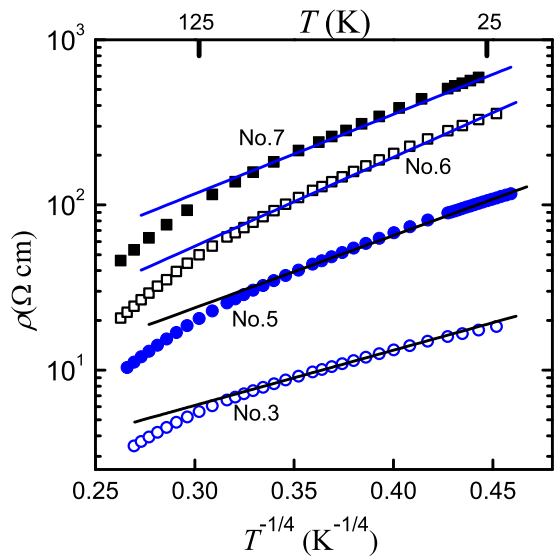}
\caption{(color online) Logarithm of resistivity as a function of $T^{-1/4}$
for four ZnO polycrystalline films, as indicated. The symbols are the
experimental data and the straight solid lines are least-squares fits to
Eq.~(\ref{Eq.(3)}).} \label{Fig.4}
\end{center}
\end{figure}

Figure~\ref{Fig.4} plots the variation of log$\rho$ with $T^{-1/4}$ for four
representative ZnO films, as indicated. The symbols are the experimental data
and the straight solid lines are the least-squares fits to the
Eq.~(\ref{Eq.(3)}) given below. Clearly, in every sample, there exists a
linear regime from $\sim$80 K down to $\sim$25 K. This observation suggests
that, in this $T$ interval, the dominate electrical conduction process is due
to the Mott VRH conduction mechanism in three dimensions: \cite{Mott79}
\begin{equation}\label{Eq.(3)}
\rho_M(T) = \rho_{M_0}\, e^{(T_M/T)^{1/4}} \,,
\end{equation}
where $\rho_{M_{0}}$ is a temperature insensitive resistivity parameter, and
$T_M$=$18/[k_B N(E_F) \xi^{3}]$ is a characteristic temperature. $N(E_F)$ is
the electronic density of states at the Fermi level, and $\xi$ is the
localization length of the relevant electronic wave function. The average
hopping distance and the average hopping energy of the conduction electrons
are, respectively, given by $\bar{R}_{\rm hop, Mott}$=$(3 \xi/8)
(T_M/T)^{1/4}$ and $\bar{W}_{\rm hop, Mott}$=$(k_BT/4)(T_M/T)^{1/4}$.
\cite{Gantmakher05} Our extracted values of $T_M$, and then calculated values
of $N(E_F)$, $\bar{R}_{\rm hop,Mott}$, and $\bar{W}_{\rm hop,Mott}$ are listed
in Table III. In order to calculate these three quantities, we have used
$\xi$$\approx$2 nm, the Bohr radius of the shallow donors in ZnO.
\cite{McCluskey09} Our fitted $\bar{R}_{\rm hop,Mott}$ values are significantly smaller than
our film thickness ($\approx$1 $\mu$m). Also, the criterion $\bar{R}_{\rm
hop,Mott}/\xi$\,$>$\,1 for Eq.~(\ref{Eq.(3)}) to be valid is satisfied in the
films Nos. 3 and 5--7. On the other hand, this criterion is not satisfied for
the films Nos. 2 and 4, because these two films are not resistive enough to
lie sufficiently deep on the insulating side of the metal-insulator
transition. \cite{Huang10} Furthermore, our extracted values of $\bar{W}_{\rm
hop,Mott}$(40\,K)$\approx$1--4 meV are somewhat smaller than the corresponding
values reported in Ref. \onlinecite{Huang10}. This small decrease in $\bar{W}_{\rm
hop,Mott}$ by $\sim$35\% from that in Ref. \onlinecite{Huang10} can be readily
understood as arising from the increased shallow donor concentrations in these
films as a result of thermal cycling/annealing. Obviously, on average, a higher donor
concentration will lead to a lower thermal activation energy involved in the
VRH conduction process.

Our observation of the Mott VRH conduction in the $T$ interval $\sim$20--80 K
in this work demonstrates that this carrier transport process in the ZnO
material is generic and robust, while the values of the relevant parameters
may differ somewhat from run to run, because the amounts of donor
concentration could have varied. At even lower measurement temperatures
($\lesssim$20 K), a crossover from the Mott VRH conduction process to the
Efros-Shklovskii VRH conduction process \cite{ES84} should be expected, as has
previously been reported in Ref. \onlinecite{Huang10}. Unfortunately, the base
temperature of our closed-cycle refrigerator does not allow $\rho (T)$
measurements down to sufficiently low $T$ to illustrate this latter VRH
conduction mechanism in the present work.

\subsubsection{Estimate of shallow donor concentration}

While the shallow donor concentration, $n_D$, in the ZnO material is a very
important quantity, it is difficult to measure directly. However, we may
estimate this quantity from the extracted $\rho_2$ value which characterizes
the $E_2$-conduction channel in Eqs.~(\ref{Eq.(1)}) and (\ref{Eq.(2)}). At
sufficiently high temperatures (in practice, above a few hundreds K), the
shallow donors should have essentially all been ionized. These excited
electrons would move in the conduction band and be in response to an
externally applied electric field. The resulted electrical conductivity from
this $E_2$-conduction channel can approximately be expressed by
$\rho_2^{-1}$=$(n_De^2 \tau_e )/m^\ast$=$n_De \mu_e$, where $\tau_e$ is the
electron elastic mean free time, $m^\ast$ is the effective electron mass, and
$\mu_e$ is the electron mobility. Typically, the value of $\mu_e$ in ZnO at
$T$ above a few hundreds K can be extrapolated to a magnitude of $\sim$10
cm$^2$/V\,s. \cite{Ellmer08} In the case of our as-grown single crystal, Table
I lists a value $\rho_2$\,$\sim$\,1$\times$$10^4$ $\Omega$ cm. Then, we may
estimate $n_D$=$(\rho_2 e \mu_e)^{-1}$$\sim$$10^{14}$ cm$^{-3}$. In the case
of our polycrystalline films, Table II lists a value $\rho_2$\,$\sim$\,1
$\Omega$ cm. Then, we may estimate $n_D$=$(\rho_2 e
\mu_e)^{-1}$$\sim$$10^{18}$ cm$^{-3}$. These inferred shallow donor
concentrations in the single crystal and in the polycrystalline films,
respectively, are in good accord with the accepted corresponding values
reported in the literature. \cite{Look03,Ellmer08} This degree of agreement
provides a self-consistency check of our data analyses and the interpretation
of the charge transport processes in ZnO in this section.

\section{Conclusion and Summary}

We have investigated the electrical conduction properties of ZnO single
crystal and polycrystalline films in the wide temperature range 20--500 K. In
the case of single crystal, we found that two types of thermal activation
conduction processes dominate the carrier transport at $T$$\gtrsim$110 K. As a
result, a group of shallow donors and a group of intermediately deep donors,
together with their individual ionization energies, have been inferred. In
particular, we observed that the shallow donor concentration is markedly
affected by thermal cycling in vacuum, strongly suggesting the important
role of oxygen vacancies as shallow donors in the ZnO material. In the case of
oxygen deficient polycrystalline films, additional conduction processes due to
the nearest-neighbor-hopping conduction and the Mott variable-range-hopping
conduction mechanisms are observed. These two additional conduction possesses
originate from the existence of far more numerous amounts of shallow donors
(oxygen vacancies) in polycrystal films, which led to the formation of a
partially filled impurity band. Taken together with our previous work (Ref.
\onlinecite{Huang10}), this present study provides a fairly complete picture of the
overall electrical conduction processes in the ZnO materials.

{\em A summary of the overall electrical conduction processes in the ZnO materials:} We would like to give a brief summary of the rich electrical conduction
processes in the intensively studied, scientifically and technologically alluring ZnO materials. As the temperature progressively increases from
liquid-hilum temperatures up to 500 K, one expects to see, one process by one process, (1) the
Efros-Shklovskii variable-range hopping conduction, (2) the Mott
variable-range hopping conduction, (3) the nearest-neighbor-hopping
conduction, (4) the thermal activation conduction from the shallow donor
levels, and finally (5) the thermal activation conduction from the
intermediately deep donors levels. The conduction processes (1)--(3) take
place in the shallow donor impurity band, while the conduction processes (4)
and (5) take place in the conduction band. These conduction mechanisms are
generic and robust in unintentionally doped n-type ZnO materials which lie on
the insulating side of the metal-insulator transition. \cite{split} However,
the temperature interval within which each conduction mechanism is found and
the extracted value of the associated parameter may vary more or less from
experiment to experiment. Such variations can readily arise from the fact that
different experiments usually involve differing amounts of donors and
differing donor level distributions in the samples.

\acknowledgments

The authors are grateful to S. P. Chiu for experimental assistance and
valuable discussion. This work was supported by the Taiwan National Science
Council through Grant No. NSC 99-2120-M-009-001 and the MOE ATU Program
(J.J.L.), and by the Key Project of Chinese Ministry of Education through
Grant No. 109042 and NSF of Tianjin City through Grant No. 10JCYBJC02400
(Z.Q.L.).


\begin{thebibliography}{100}
\bibitem{backflow} Well below room temperatures, the closed-cycle refrigerator served as a cryopump. The diffusion pump was therefore disconnected to prevent a backflow of its oil vapor into the refrigerator.
\bibitem{rich} The intermediately deep donor concentration should be much
higher than the shallow donor concentration. This can be seen from the fact
that, once $T$ is sufficiently high and the intermediately deep donors are
excited, the $E_1$-conduction channel dominates the sample conductance. Notice
that in Table~\ref{Table I} the $\rho_1$ values are 5 to 6 orders of magnitude
lower than the corresponding $\rho_2$ values. An estimate of the shallow donor
concentration is given in subsection IIIB.3.
\bibitem{disperse} Strictly speaking, the $E_2$ values are sightly reduced from
$\approx$29 to $\approx$25--26 meV after repeated thermal cycling. This can
be ascribed to a slight increase in the donor concentration and thus an associated weak dispersion of the
shallow impurity levels.
\bibitem{thermionic} In case the grain boundary in a polycrystalline
semiconductor film forms a depletion region, the thermionic emission effect on
the electrical conduction might need to be taken into consideration. See, for
example, Refs. \onlinecite{Heluani06}  and \onlinecite{Tsurumi99}. In this
study, our films are single-phased and possess a good $c$-axis texture (Ref.
\onlinecite{Huang10}), and thus the grain boundary effect does not play a
notable role.
\bibitem{split} If a sample lies sufficiently close to the metal-insulator
transition, the ``split-impurity-band" conduction process may take place, as
has recently been observed in relatively heavily and natively doped,
single-crystalline ZnO nanowires (Refs. \onlinecite{Chiu09} and
\onlinecite{Tsai10}). In this case, the shallow impurity band splits into a
lower subband and an upper subband as a consequence of the presence of
excessive randomness and strong Coulomb interactions, see Refs.
\onlinecite{Mott79} and \onlinecite{ES84}.
\end{thebibliography}

\end{document}